\documentclass[sigconf,noacm]{acmart}

\usepackage{amsmath,amsfonts}
\usepackage{pifont}
\usepackage{algorithm}
\usepackage[noend]{algpseudocode}
\usepackage{array}
\usepackage[labelfont=scriptsize,textfont=scriptsize]{subfig}
\usepackage{textcomp}

\usepackage{enumitem}
\usepackage{stfloats}
\usepackage{tcolorbox}
\usepackage{xurl}
\usepackage{multirow}
\usepackage{threeparttable}
\usepackage{booktabs}
\usepackage{multirow}
\usepackage{balance}
\usepackage{verbatim}
\usepackage{graphicx}
\usepackage{xspace}	

\newcommand{\sys}{\textsc{PonziSleuth}\xspace}
\makeatletter
\def\hlinewd#1{%
	\noalign{\ifnum0=`}\fi\hrule \@height #1 \futurelet
	\reserved@a\@xhline}
\makeatother

\usepackage{hyperref}
\hypersetup{colorlinks=true, linkcolor=blue, citecolor=blue, filecolor=blue, urlcolor=blue,  pdfpagemode=FullScreen}

\usepackage{listings}
\usepackage{xcolor}
\usepackage{listing-solidity}
\definecolor{highlight}{rgb}{1,1,0.6} 

\setcopyright{none}
\renewcommand\footnotetextcopyrightpermission[1]{} 
\usepackage{bbding}

\begin{document}

\title{Semantic Sleuth: Identifying Ponzi Contracts via Large Language Models}

\author{Cong Wu, Jing Chen\Envelope, Ziwei Wang, Ruichao Liang, Ruiying Du\\
		\emph{School of Cyber Science and Engineering, Wuhan University, China}\\
		\texttt{\{cnacwu,chenjing,t4stek1ng,liangruichao,duraying\}@whu.edu.cn}}

\begin{abstract}
	Smart contracts, self-executing agreements directly encoded in code, are fundamental to blockchain technology, especially in decentralized finance (DeFi) and Web3. However, the rise of Ponzi schemes in smart contracts poses significant risks, leading to substantial financial losses and eroding trust in blockchain systems. Existing detection methods, such as PonziGuard, depend on large amounts of labeled data and struggle to identify unseen Ponzi schemes, limiting their reliability and generalizability. In contrast, we introduce \sys, the first LLM-driven approach for detecting Ponzi smart contracts, which requires no labeled training data. \sys utilizes advanced language understanding capabilities of LLMs to analyze smart contract source code through a novel two-step zero-shot chain-of-thought prompting technique. Our extensive evaluation on benchmark datasets and real-world contracts demonstrates that \sys delivers comparable, and often superior, performance without the extensive data requirements, achieving a balanced detection accuracy of 96.06\% with GPT-3.5-turbo, 93.91\% with LLAMA3, and 94.27\% with Mistral. In real-world detection, \sys successfully identified 15 new Ponzi schemes from 4,597 contracts verified by Etherscan in March 2024, with a false negative rate of 0\% and a false positive rate of 0.29\%. These results highlight \sys's capability to detect diverse and novel Ponzi schemes, marking a significant advancement in leveraging LLMs for enhancing blockchain security and mitigating financial scams.

\end{abstract}
\pagestyle{plain}
\maketitle

\section{Introduction}
Smart contracts, self-executing contracts with the terms of the agreement directly written into code, are foundational to decentralized finance and Web3~\cite{sun2023panda,liang2024vulseye,liang2024towards,wu2022TokenScout}. However, their widespread adoption has also facilitated the emergence of Ponzi schemes, which are fraudulent investment operations where returns to earlier investors are paid using the capital from newer investors~\cite{chen2024ponzifinder,liang2024ponziguard}. These schemes exploit the transparency and automation of smart contracts by promising high returns with minimal risk, thereby attracting new investors. The funds from these new investors are used to pay earlier investors, creating a false impression of a profitable venture. As the influx of new investors diminishes, the scheme inevitably collapses, resulting in significant financial losses for most participants.
Ponzi schemes erode trust in DeFi and Web3 ecosystems, cause severe economic damage, and hinder broader adoption~\cite{lou2020ponzi}.

Several methods have been developed to detect Ponzi contracts on Ethereum, which can be categorized into three main types. The first type uses bytecode or opcodes from smart contracts to train classifiers or perform static analysis~\cite{fan2021spsd,chen2018detecting}. These methods analyze the low-level instructions executed by the Ethereum virtual machine to identify patterns that suggest Ponzi schemes. The second type examines the transaction behavior of smart contracts, looking at the sequence and characteristics of transactions to spot suspicious financial activities~\cite{yu2021ponzi,cai2023ponzi}. The third type combines both opcode features and account features to train detection models, using a wider range of data to improve accuracy~\cite{lou2020ponzi,liang2024ponziguard}. Additionally, there are rule-based approaches that rely on expert knowledge of Ponzi schemes, creating predefined rules to identify fraudulent contracts based on known patterns~\cite{chen2021sadponzi,sun2020early}.

\textbf{Motivation.}
Despite significant advancements, existing detection methods face critical limitations. They depend heavily on large volumes of labeled training data, which is often challenging and costly to obtain~\cite{chen2021sadponzi,liang2024ponziguard,lu2024sourcep}. Additionally, these methods rely on static information, which does not adequately capture the dynamic and evolving nature of Ponzi schemes, leading to poor reliability and limited interpretability. Furthermore, they struggle to detect new and previously unseen Ponzi contracts due to their inability to understand the behavioral semantics embedded in the contract code~\cite{sun2020early}. In contrast, large language models (LLMs) have shown significant promise in understanding complex code semantics and human language~\cite{chiang2023can,chen2021evaluating,xu2023lmpa,0x2a53f4,lin2023pushing,fang2024automated,lin2024splitlora}. LLMs, including GPT-3, LLAMA, and their variants, are advanced neural networks trained on vast amounts of text data, enabling them to comprehend context, identify patterns, and generate accurate analyses~\cite{lin2024fedsn,lin2024efficient,fang2024ic3m,lin2024adaptsfl}. 
This capability makes them particularly suited for interpreting and detecting diverse Ponzi schemes, even those not encountered during training. Therefore, our research question is: \textit{Can large language models be designed to efficiently and robustly identify Ponzi contracts, including previously unseen ones, using minimal labeled data?}

\textbf{\sys.} In this paper, we propose \sys, the first LLM-driven approach  for detecting Ponzi contracts. \sys leverages LLMs to analyze source codes of contracts, enabling efficient and robust identification of Ponzi schemes with minimal labeled data. \sys combines zero-shot chain-of-thought (Zero-shot-CoT) prompting, static taint analysis, and automated code slicing, allowing \sys to grasp the complex behavioral semantics of Ponzi contracts that traditional methods miss.

\sys operates in three main phases. First, in the preparation phase, the source code of a smart contract is compiled to generate an abstract syntax tree (AST), which is then parsed into an intermediate representation (IR) and a structured contract object. Second, in the contract analysis and slicing phase, static taint analysis traces the flow of funds within the contract, creating a taint propagation graph that visualizes data flow. Relevant code slices are extracted to reduce the input size for the LLM. Finally, in the LLM detection phase, the contract slices and taint propagation graph are used to generate high-quality prompts for the LLM, which then analyzes the contract to identify potential Ponzi schemes.
By uncovering the semantic behavior of Ponzi contracts, \sys overcomes the limitations of traditional static analysis and rule-based detection methods. It addresses the challenge of relying on large amounts of training data and enhances the ability to recognize previously unseen Ponzi contracts.

Our evaluation shows that \sys performs robustly across different LLMs, achieving a BAC of 96.06\% with GPT-3.5-turbo, 93.91\% with LLAMA3, and 94.27\% with Mistral, demonstrating its capability in detecting both known and previously unseen Ponzi contracts. Unlike PonziGuard, which relies heavily on labeled data and may be prone to overfitting, \sys offers a more generalized approach, effectively identifying 15 new Ponzi schemes from 4,597 contracts verified by Etherscan between March 14-24, 2024, with a false negative rate (FNR) of 0\% and a false positive rate (FPR) of 0.29\%. We responsibly disclosed the detected Ponzi contracts to the Web3 security community. Additionally, \sys demonstrates impressive efficiency, processing each contract in an average of 5.52 seconds at a cost of \$0.0027 per contract using 2601.3 tokens per detection. \sys has been deployed for real-time monitoring of Ponzi contracts, providing a powerful and innovative tool for detecting fraud in smart contracts without relying on extensive labeled data.

In summary, we make the following contributions.
\begin{itemize}
	
	\item {To our knowledge, \sys is the first LLM-driven method for detecting Ponzi smart contracts, requiring none labeled training data.}
	
	\item We introduce a novel two-stage Zero-shot-CoT method in \sys, enhancing LLMs with static taint analysis and automated code slicing to efficiently detect Ponzi schemes, understanding complex fraud patterns, tracing tainted data, and focusing on relevant code segments without extensive retraining.
	
	\item {We perform comprehensive evaluations of \sys across various settings, including extensive comparisons with existing methods, real-world detection, overhead evaluation.}
	Results indicate that \sys is effective in detecting unseen Ponzi contracts without relying on labeled data, offering a significant advantage over existing approaches.
\end{itemize}

The data and codebase of \sys can be obtained at the repository: {https://github.com/tasteking/PonziSleuth-ASE24}.

\section{Background}
Ponzi schemes are fraudulent investment operations where returns to earlier investors are paid using the capital from newer investors, rather than from profit earned by the operation~\cite{chen2021sadponzi,lu2024sourcep,sun2020early,liang2024ponziguard,bartoletti2020dissecting,lou2020ponzi}. These schemes promise high returns with little risk, enticing many investors. A typical Ponzi contract, such as 0x2a...1e9~\cite{0x2a53f4}, shown in Listing~\ref{code:exp1}, attracts greedy investors by promising high returns.
Whenever an investor calls the \texttt{enter} function to invest, the function appends the investor's address to the end of the investor list and then iteratively pays previous investors twice the amount they invested, until the contract's balance is insufficient to continue these payments. Although the first few investors may receive the promised returns, this unsustainable model eventually collapses, leaving later investors with significant losses.

\begin{figure}[!h]
	\centering
	\includegraphics[width = \linewidth]{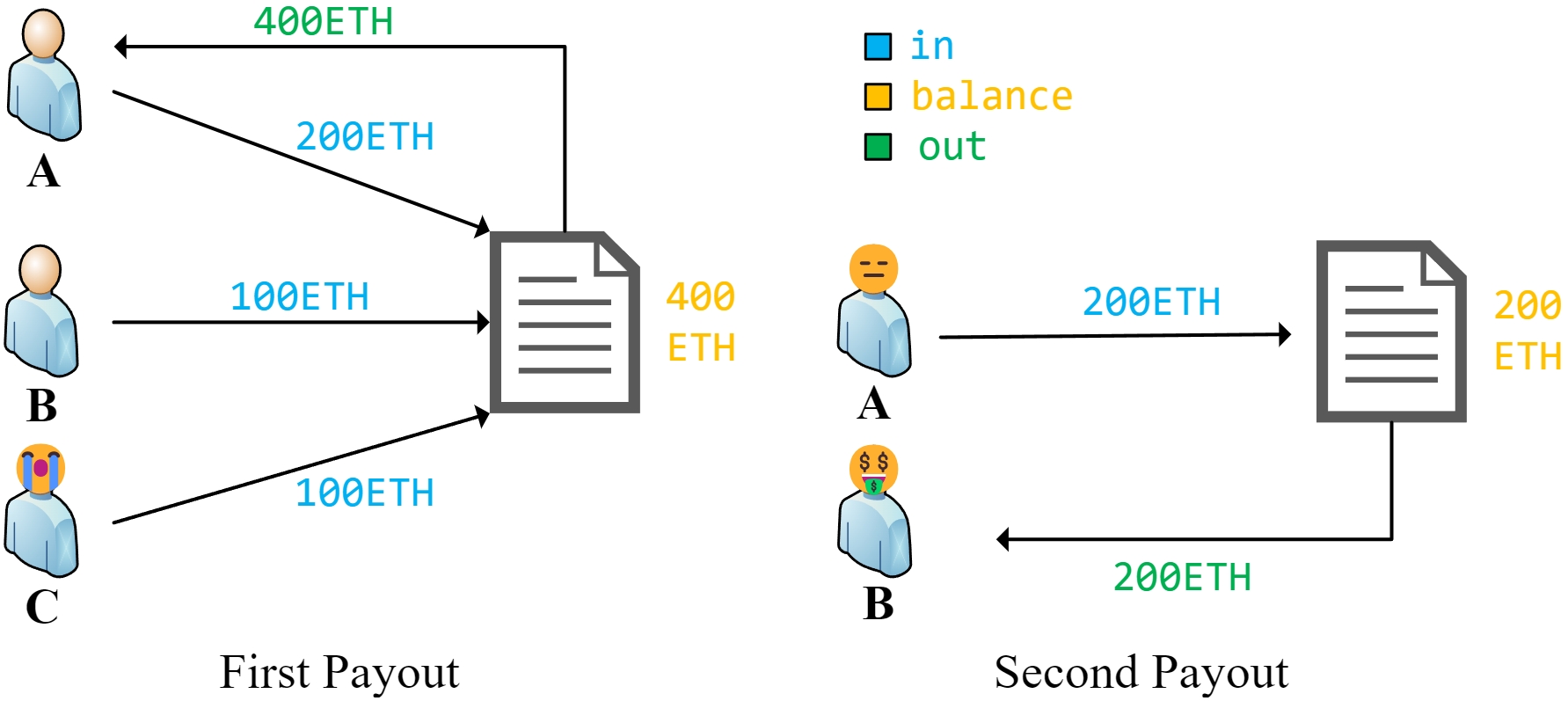}
	\caption{Illustration of payout mechanism in Ponzi contract}
	\label{fig:exp}
\end{figure}

Figure~\ref{fig:exp} visualizes the payout mechanism of a Ponzi contract. In the first payout phase, the contract collects a total of 400 ETH from investors A, B, and C and pays out 400 ETH to A, the first investor, leaving a balance of 400 ETH. In the second phase, 200 ETH is collected from A and paid to B, the second investor, reducing the balance to 200 ETH. It illustrates the contract's dependence on continuous new investments to pay earlier investors. When new investments cease, the contract balance becomes insufficient to continue payouts, leading to the scheme's collapse and losses for the last investors.

To detect such fraudulent schemes, our aim is to leverage the advanced reasoning capabilities of LLMs for semantic analysis, making them powerful tools for uncovering sophisticated frauds and addressing the limitations of traditional static analysis and rule-based detection methods. We aim to utilize a novel zero-shot chain-of-thought prompting technique that enables the LLM to analyze the smart contract's logic and identify patterns indicative of Ponzi schemes. Our motivation is to create an efficient and accurate detection system that does not rely on extensive retraining. By structuring the prompts in a two-step manner, i.e., understanding the contract's functionality and detecting Ponzi characteristics, we seek to exploit the LLM's ability to provide detailed, context-aware analyses.

\begin{lstlisting}[caption={Code snippet of contract 0x2a...1e9},label={code:exp1} ] 
	function enter() {
		uint amount;
		amount = msg.value;
		uint idx = persons.length; 
		persons.length += 1; 
		persons[idx].etherAddress = msg.sender;
		persons[idx].amount = amount;
		balance += amount;
		while (balance >= persons[payoutIdx].amount * 2) { 
			uint transactionAmount = persons[payoutIdx].amount * 2; 
			persons[payoutIdx].etherAddress.send(transactionAmount); 
			balance -= transactionAmount;
			payoutIdx += 1;	}}
\end{lstlisting}

\section{Overview}
In this section, we brief threat model regarding Ponzi contracts and present the overview of \sys.
\subsection{Threat Model}

\textbf{Understanding Ponzi contracts.}
Ponzi contracts are fraudulent schemes implemented as smart contracts on blockchain platforms like Ethereum. They promise high returns with little risk to attract investors~\cite{galletta2024explainable}. Instead of generating profits from legitimate business activities, Ponzi contracts use funds from new investors to pay returns to earlier investors, creating an illusion of profitability. This deceptive operation is embedded in code's semantic behavior, where the logic of fund redistribution is designed to perpetuate the scheme as long as new investments continue to flow.

\textbf{Operational phases of Ponzi contracts.}
Ponzi contracts typically operate in several phases, each reflected in the contract's code semantics. In the initial phase, the attacker creates a smart contract promising high returns on investments, often incorporating seemingly legitimate business logic to gain credibility. During the growth phase, early investors receive returns from new investors' funds, with the contract's code managing these transactions to maintain the illusion of profitability. As the scheme grows, it enters the saturation phase, where attracting new investors becomes increasingly difficult, and the contract's logic struggles to fulfill its promises. Finally, in the collapse phase, the inflow of new investments slows or stops, causing the scheme to fail and leading to significant financial losses for most participants, except for the initial fraudsters who designed the scheme to benefit themselves.

\textbf{Attacker strategies and code semantics.}
Attackers use various strategies to attract victims to their Ponzi schemes, which are closely tied to the semantic behavior of the smart contract code. High return promises are embedded in the contract's promotional materials and code, enticing victims with the allure of substantial profits. Early payouts are programmed into the contract to build credibility and attract more victims, leveraging initial successes to foster trust. Social engineering tactics, such as persuasive marketing and endorsements, are supported by complex and obfuscated contract code, making it difficult for victims to identify the fraud. Referral bonuses coded into the contract encourage existing investors to recruit new participants, perpetuating the scheme. These tactics not only exploit human vulnerabilities but also pose serious threats to system security by undermining the integrity of blockchain ecosystems.
Understanding these mechanisms and their semantic implications within the code is crucial for developing effective detection methods. By analyzing these patterns and behaviors, \sys leverages the advanced capabilities of LLMs to provide robust and efficient detection of Ponzi contracts.

\begin{figure*}[!t]
	\centering
	\includegraphics[width = .85\linewidth]{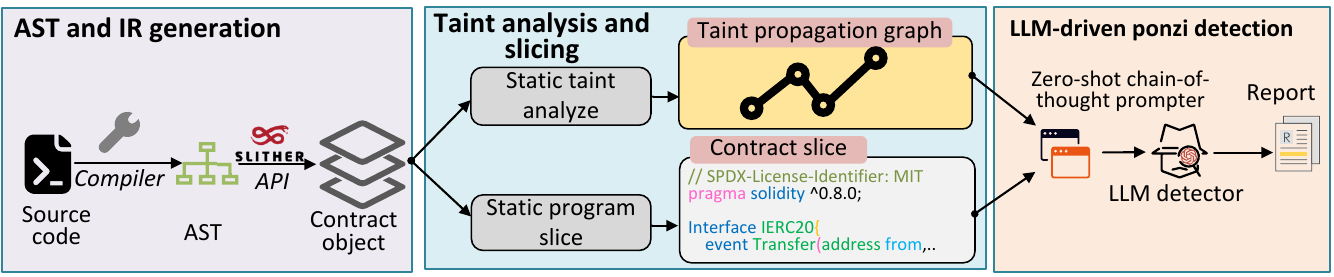}
	\caption{Workflow of \sys}
	\label{fig:overview}
\end{figure*}

\subsection{Overview of \sys}

As illustrated in Figure~\ref{fig:overview}, \sys consists of three phases: AST and IR Generation, Taint Analysis and Slicing, and LLM-driven Ponzi Detection. In the first phase, the smart contract's source code is compiled and parsed to generate an AST and IR, providing a structured view of the contract's code. This representation breaks down complex code into manageable components. The second phase involves static taint analysis to trace the flow of tainted data within the contract, creating taint propagation graphs that visualize data movement through the code. By isolating relevant code slices, we reduce the input size for the LLM, improving analysis efficiency. In the final phase, these contract slices and taint graphs generate high-quality prompts for the LLM. The LLM uses a Zero-shot-CoT prompting technique to reason about the code and identify potential Ponzi schemes. This comprehensive analysis results in a detailed report highlighting the findings and potential risks of the analyzed smart contract, providing clear insights and actionable information.

\section{Design of \sys}

In this section, we detail the design of \sys.

\subsection{AST and IR Generation}
In the AST and IR Generation phase, the smart contract's source code is compiled into an AST using the Solidity compiler. The AST provides a hierarchical representation of the contract's syntax, breaking it down into fundamental components like functions and variables. This structured format simplifies the analysis of the code. The AST is then parsed to generate an IR, which abstracts the contract's logic into a more analyzable form, encapsulating both syntactic and semantic details. This process transforms complex smart contract code into structured formats that facilitate accurate taint analysis and code slicing, setting a solid foundation for detecting Ponzi schemes with precision.

\subsection{Taint Analysis and Slicing}
This step aims to trace the flow of funds within the smart contract to identify potential Ponzi scheme behaviors and isolate relevant parts of the contract code for efficient analysis by the LLM. The intuition behind this phase is that by following the flow of tainted data, such as ether sent to the contract and the sender's address, we can detect patterns indicative of Ponzi schemes. Providing this detailed information to the LLM helps it understand the contract's logic more accurately.

\textbf{Taint analysis.}
To achieve this, we start by identifying the sources of tainted data, such as the ether sent to the contract and the sender's address. Static taint analysis is performed to track how these tainted variables propagate through the contract. Specifically, we model the contract as a hypernode graph \( H_c = (G_c, N_c, E_c) \), where \( G_c \) is the graph identifier, \( N_c \) is the set of nodes, and \( E_c \) is the set of edges.

(i) \emph{Graph representation}: For a given hypernode \( G \), its graph representation includes nodes \( N \) and edges \( E \).
\[
\text{graph}(G) = (G, N, E)
\]

(ii) \emph{Node identification}: Nodes can be basic nodes or other hypernodes, representing contract components like functions or variables.
\[
\text{nodes}(G) = \{ n \mid n \in \mathbf{N} \cap N \} \cup \{ G' \mid G' \in \mathbf{G} \cap N \}
\]

(iii) \emph{Edge construction}: Edges are created between nodes to represent data flow.
\begin{align}
	\nonumber   \text{edges}(G) = & \{ (n_1, n_2) \mid (n_1, n_2) \in E \}              \\
	\nonumber                     & \cup \{ (n, G') \mid (n, \text{graph}(G')) \in E \} \\
	& \cup \{ (G', n) \mid (\text{graph}(G'), n) \in E \}
\end{align}

The taint propagation is visualized using a taint propagation graph, \( H_s = (G_s, N_s, E_s) \), which illustrates the data flow within the contract. Initially, the subgraph \( N_s \) is set to the taint sources \( \{ n_{msg.sender}, n_{msg.value} \} \). The algorithm iteratively adds nodes and edges that are affected by the tainted data until no further propagation occurs. Algorithm~\ref{alg:taint} details the Taint Propagation Algorithm.

Note that this may lead to over-tainting, where all potential branches are analyzed, even if they are not executable. Additionally, it does not account for modifications by non-tainted variables, resulting in all tainted state variables remaining tainted. Despite this imprecision, the analysis effectively highlights suspicious code and variables, helping identify potential Ponzi scheme behavior. This thorough tracing ensures that no critical paths are missed, making it a robust method for detecting fraudulent contracts.

\textbf{Contract slicing.}
Once the taint propagation graph is generated, the next step is contract slicing to extract the relevant code segments. This process ensures computational efficiency and cost-effectiveness when feeding the code into the LLM. By slicing the contract at the function level, we preserve the logical integrity of the extracted segments, focusing on the flow of funds, which is crucial for detecting Ponzi scheme behaviors. This method ensures that only the essential parts of the contract are analyzed, reducing unnecessary computational overhead and costs while maintaining the contract's logical structure. It allows the LLM to concentrate on critical aspects related to fund flow and potential Ponzi scheme indicators, thereby enhancing the accuracy and efficiency of the detection process.

(i) \emph{Function level slicing.}
The goal is to extract code segments that interact with tainted data, ensuring that the LLM analyzes only the relevant parts of the contract. Let \( F = \{f_1, f_2, \ldots, f_n\} \) be the set of functions in the contract. We define \( S \subseteq F \) as the subset of functions that interact with tainted data:
\[
S = \{ f \mid f \in F \text{ and } \exists v \in N_s, \text{ where } v \text{ is used in } f \}
\]
This subset \( S \) contains all functions that either use or modify tainted variables identified during the taint analysis phase.

(ii) \emph{Combining slices.}
To ensure a comprehensive understanding of the contract's logic, it is necessary to combine relevant slices. The combined slice \( C \) is the union of all individual function slices:
\[
C = \bigcup_{f \in S} \text{slice}(f)
\]
where \( \text{slice}(f) \) represents the code segment corresponding to function \( f \). By doing so, we ensure that the LLM receives a coherent and complete representation of the fund flow within the contract.

\begin{algorithm}[!t]
	\small
	\caption{Taint Propagation Algorithm: TPA()}
	\begin{algorithmic}[1]
		\Statex \textbf{Input:} Contract hypernodes \( H_c = (G_c, N_c, E_c) \)
		\Statex \textbf{Output:} Taint propagation subgraph \( H_s = (G_s, N_s, E_s) \)
		\State Initialize \( \text{coverage} \gets \{\} \)
		\State Initialize \( H_s' \gets H_s \)
		\While{true}
		\State \( H_s' \gets H_s \)
		\For{$G \in \text{nodes}(G_c) \cap \mathbf{G}$}
		\If{$G \in \text{coverage}$}
		\State \textbf{continue}
		\EndIf
		\For{$(n_0, n_1) \in \text{edges}(G)$}
		\If{$n_0 \in N_s$}
		\If{$n_1 \in \mathbf{N}$}
		\State $N_s \gets N_s \cup \{n_1\}$
		\State $E_s \gets E_s \cup \{(n_0, n_1)\}$
		\ElsIf{$n_1 \in \mathbf{G}$}
		\State $H_s \gets H_s \cup \text{TPA}(n_1)$
		\State $\text{coverage} \gets \text{coverage} \cup \{n_1\}$
		\EndIf
		\EndIf
		\EndFor
		\State $\text{coverage} \gets \text{coverage} \cup \{G\}$
		\EndFor
		\If{$H_s = H_s'$}
		\State \textbf{break}
		\EndIf
		\EndWhile
		\State \Return $H_s$
	\end{algorithmic}
	\label{alg:taint}
\end{algorithm}

\subsection{LLM-Driven Ponzi Detection}

The aim of this phase is to leverage the advanced reasoning capabilities of LLMs to detect Ponzi schemes in smart contracts. By using carefully designed prompts, we enhance the LLM's ability to analyze the contract's logic and identify suspicious patterns indicative of Ponzi schemes. Prior research has shown that prompt engineering can significantly improve LLM performance, particularly in complex tasks like Ponzi scheme detection where the diversity of contract types poses a challenge. Few-shot prompts often suffer from limited context, and one-shot prompts may fail to encompass the variety of Ponzi schemes. Therefore, we employ Zero-shot-CoT prompting, which enables the LLM to perform insightful analyses without extensive retraining, leveraging its reasoning abilities to detect fraudulent patterns effectively.
We design a two-step Zero-shot-CoT prompting technique: analyzing contract logic and detecting ponzi schemes.

\textbf{Analyzing contract logic.}
In the first step, we organize the code slices and taint propagation graphs into prompts designed to guide the LLM in analyzing the logic of individual functions and the overall contract. These prompts include specific questions and instructions to help the LLM decompose the code, understand the flow of funds, and identify any anomalies. The LLM then outputs a detailed analysis report for each function and the contract as a whole.

Let $\text{Code}_i$ represent the $i$-th code slice and $\text{TaintGraph}_i$ represent the corresponding taint propagation graph. The prompt for the LLM can be formulated as:
\[
\text{Prompt}_1 = \{ \text{Code}_i, \text{TaintGraph}_i \} \rightarrow \text{LLM}(\text{Prompt}_1) = \text{Analysis}_i
\]
where $\text{Analysis}_i$ is the output analysis report for the $i$-th function or contract slice.

\textbf{Detecting Ponzi contracting.}
In the second step, the analysis results from the first step are combined with the formal definition of a Ponzi scheme. This combined prompt is then fed to the LLM to determine if the analyzed contract exhibits Ponzi scheme characteristics. The LLM uses its understanding of the contract's behavior, along with the provided definition, to make an informed decision. The outcome is a detection report indicating whether the contract is likely to be a Ponzi scheme.

Let $\text{Def}_{\text{Ponzi}}$ represent the formal definition of a Ponzi scheme. The combined prompt for the LLM can be formulated as:
\[
\text{Prompt}_2 = \{ \text{Analysis}_i, \text{Def}_{\text{Ponzi}} \} \rightarrow \text{LLM}(\text{Prompt}_2) = \text{Detection}_i
\]
where $\text{Detection}_i$ is the result indicating whether the $i$-th function or contract slice exhibits Ponzi scheme characteristics.


\section{Implementation Details}
We use the Solc~\cite{solc} Python API to compile smart contracts and generate their ASTs. The Slither-analyze~~\cite{slither} then parses these ASTs to produce IRs and structured contract objects, providing essential information for further analysis. Using these contract objects, we construct hypernodes to represent the contracts and generate taint propagation subgraphs through the taint analysis module. For visualization, the subgraph is converted to a regular graph in Graphviz format, merging contract state variables. The slicing module uses the node-to-source code mapping created during hypernode construction to extract relevant code slices, ensuring a focused and efficient analysis. The detection module then uses the taint propagation graphs and code slices to automatically generate prompts for various LLMs, which are fed into models like
GPT-3.5-turbo~\cite{gpt35}, LLAMA2-7b~\cite{Llama2}, LLAMA3-8b~\cite{Llama3}, and Mistral-7b~\cite{Mistral}.
GPT-3.5-turbo interacts with the OpenAI Python API, while the other open-source models run locally using the Ollama framework, all configured with a temperature setting of 0 for deterministic outputs.

\begin{figure}[!h]
	\centering
	\includegraphics[width = \linewidth]{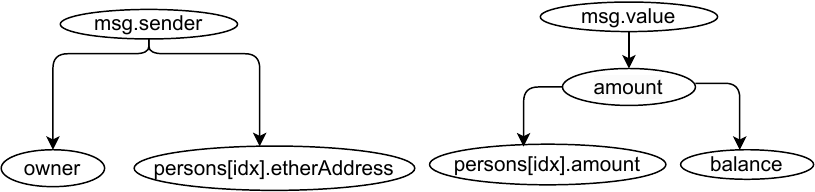}
	\caption{Example of contract 0x2a...1e9's taint propagation graph}
	\label{fig:tpexp}
\end{figure}

This module maps nodes to their corresponding source code segments during the construction of hypernodes. This allows us to extract relevant code slices at the function level, ensuring the logical integrity of the contract is maintained. An example of function-level code slices is provided below:
An example of a taint propagation graph is illustrated as Figure~\ref{fig:tpexp}.

\begin{figure}[!h]
	\centering
	\includegraphics[width = \linewidth]{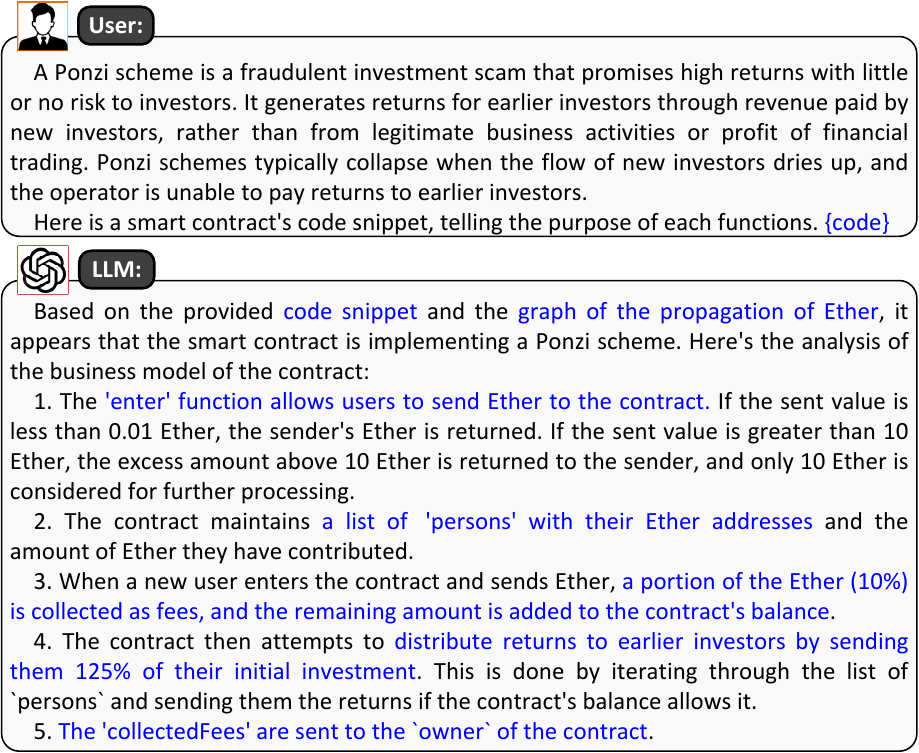}
	\caption{Example of understanding contracts}
	\label{fig:1}
\end{figure}

\begin{figure}[!h]
	\centering
	\includegraphics[width = \linewidth]{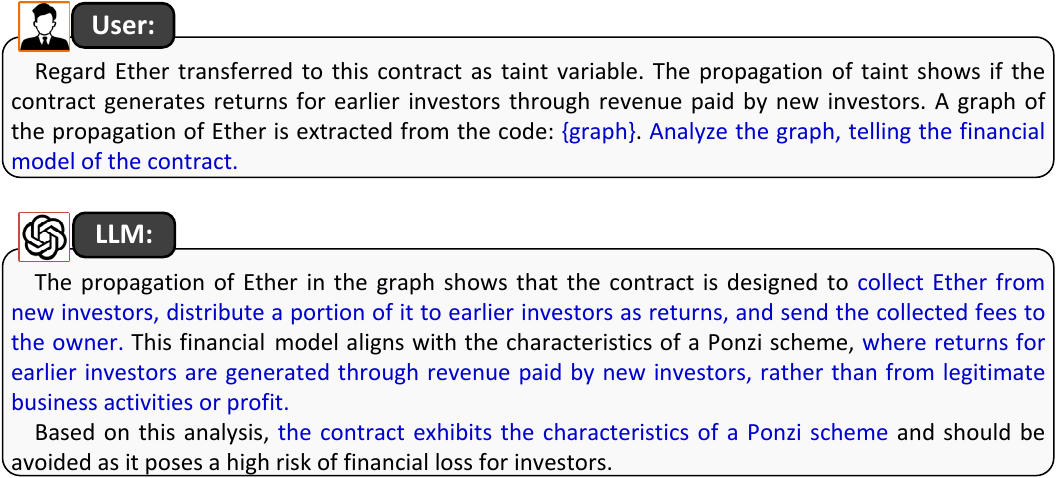}
	\caption{Example of analyzing taint propagation graph}
	\label{fig:2}
\end{figure}

\begin{figure}[!h]
	\centering
	\includegraphics[width = 0.6\linewidth]{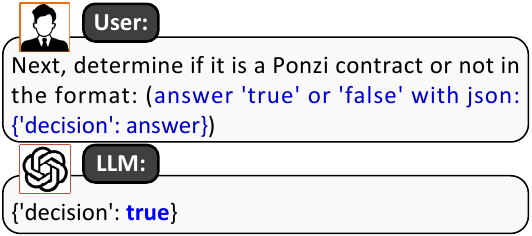}
	\caption{Example of detection results by LLM}
	\label{fig:4}
\end{figure}

Figures~\ref{fig:1}, \ref{fig:2}, and \ref{fig:4} illustrate examples under different stages of \sys's interaction with LLM for detecting Ponzi schemes in smart contracts. Figures~\ref{fig:1} shows an example of the LLM understanding the contract's purpose, where it analyzes the provided code snippet and taint propagation graph to explain the contract's business model. Figures~\ref{fig:2} demonstrates the LLM's capability to analyze a taint propagation graph, where it assesses the financial model of the contract to determine if it generates returns for earlier investors through revenue paid by new investors.
Figures~\ref{fig:4} presents the detection results, where the LLM makes a decision on whether the contract is a Ponzi scheme based on the provided information, outputting a clear `true' or `false'.

\section{Performance Evaluation}
We evaluated \sys on widely used benchmark datasets and real-world smart contracts, addressing the following research questions: \textbf{RQ1.} What is the overall performance of \sys in detecting Ponzi contracts? \textbf{RQ2.} How does \sys compare to SOTA detection methods? \textbf{RQ3.} What is the effectiveness of the individual modules in \sys (ablation study)?
\textbf{RQ4.} How well does \sys perform in detecting smart contracts compiled with different Solidity versions in a balanced dataset?
\textbf{RQ5.} How effective is \sys in detecting latest real-world Ponzi contracts?
\textbf{RQ6.} What are the computational costs of \sys under different settings?

\subsection{Experimental Setup}

\textbf{Datasets.}
We utilized several datasets to evaluate \sys's performance and robustness.
For RQ1, RQ2, RQ3, and RQ6, we used a widely-recognized benchmark dataset~\cite{bartoletti2020dissecting,liang2024ponziguard,xblock,PonziDataset,chen2021sadponzi,lu2024sourcep,sun2020early}, consisting of 139 Ponzi contracts and 1,312 non-Ponzi contracts.
These contracts, with an average length of 400 lines, are primarily from version 0.4.x, and 75\% of them have fewer than 500 lines.
For RQ4, we created a new near-balanced dataset by randomly sampling 340 Ponzi contracts and 300 non-Ponzi contracts, covering various Solidity versions from 2016 to March 2024.
For RQ5, we tested \sys's prototype on 4,597 Etherscan-verified contracts in March 14-24, 2024.

\textbf{Metrics.}
We evaluate the performance of our Ponzi contract detection method using several standard metrics: True Positive Rate (TPR), True Negative Rate (TNR), FPR, FNR, and Balanced Accuracy (BAC). TPR measures the proportion of actual Ponzi contracts correctly identified, while TNR measures the proportion of non-Ponzi contracts correctly identified. FPR and FNR indicate the rates of false positives and false negatives, respectively. BAC provides an average of TPR and TNR, offering a balanced measure of performance. Additionally, we assess the overhead in terms of average processing time, token usage, and cost, particularly for models like GPT-3.5-turbo and open-source models such as LLAMA2, LLAMA3, and Mistral.

We evaluated \sys against state-of-the-art methods, including SADPonzi~\cite{chen2021sadponzi}, PonziGuard~\cite{liang2024ponziguard}, and SourceP~\cite{lu2024sourcep}, using Solidity versions 0.4.11 to 0.8.23. The setup involved \texttt{slither-analyzer} 0.10.2, \texttt{openai} 1.23.6, and \texttt{ollama} 0.1.32, with data split into 45\% for training, 10\% for validation, and 45\% for testing. The evaluation ran on a server equipped with an Intel Xeon Gold 6133 CPU, 256GB RAM, and 3 Nvidia GeForce 4090 GPUs, using models GPT-3.5-turbo~\cite{gpt35}, LLAMA2-7b~\cite{Llama2}, LLAMA3-8b~\cite{Llama3}, and Mistral-7b~\cite{Mistral}. Each prompt was executed five times across the LLMs, with average performance reported to account for response variability and ensure robust results for \sys.

\subsection{RQ1. Overall Performance of \sys}

Table~\ref{tab:overall_p} reports the overall performance of \sys, demonstrating that \sys with GPT-3.5-turbo model achieved the highest BAC of 96.06\%, with a TPR of 96.40\% and a TNR of 95.71\%. This indicates robust detection capabilities and a strong balance between detecting Ponzi contracts and correctly identifying non-Ponzi contracts. The llama3 model also performed well, achieving a BAC of 93.91\%, TPR of 95.68\%, and TNR of 92.14\%, suggesting it as a strong open-source alternative for Ponzi contract detection. Similarly, the mistral model demonstrated effective performance with a BAC of 94.27\%, TPR of 95.68\%, and TNR of 92.86\%. In contrast, the llama2 model exhibited significantly lower performance, with a BAC of 82.58\%, TPR of 79.14\%, and TNR of 86.02\%. The high FPR of 20.86\% and FNR of 13.98\% for llama2 indicate its difficulty in accurately detecting Ponzi contracts and distinguishing them from non-Ponzi contracts. This analysis highlights the superiority of the GPT-3.5-turbo model in terms of accuracy and reliability, while also recognizing the capability of llama3 and mistral as effective open-source alternatives, though with slightly lower performance.

\begin{table}[!h]
	\centering
	\small
	\caption{Performance of \sys under different LLMs}
	\begin{tabular}{lccccc}
		\hline
		\textbf{Model}         & \textbf{TPR} & \textbf{TNR} & \textbf{FNR} & \textbf{FPR} & \textbf{BAC} \\
		\hline
		\textbf{GPT-3.5-turbo} & 96.40\%      & 95.71\%      & 3.60\%       & 4.29\%       & 96.06\%      \\
		\textbf{LLAMA2}        & 79.14\%      & 86.02\%      & 20.86\%      & 13.98\%      & 82.58\%      \\
		\textbf{LLAMA3}        & 95.68\%      & 92.14\%      & 4.32\%       & 7.86\%       & 93.91\%      \\
		\textbf{Mistral}       & 95.68\%      & 92.86\%      & 4.32\%       & 7.14\%       & 94.27\%      \\
		\hline
	\end{tabular}
	\label{tab:overall_p}
\end{table}

\subsection{RQ2. Comparing with Existing Methods}
We evaluated \sys under the GPT-3.5-turbo model against state-of-the-art Ponzi detection tools, including SADPonzi~\cite{chen2021sadponzi}, PonziGuard~\cite{liang2024ponziguard}, and SourceP~\cite{lu2024sourcep}, using varying ratios of training data. Table~\ref{tab:comp} presents the performance comparison across different methods. \sys, which does not require labeled training data, achieved a balanced accuracy (BAC) of 96.06\%, with a TPR of 96.40\% and a TNR of 95.71\%. This performance is notable given that PonziGuard, which relies on labeled training data, achieved a slightly higher BAC of 98.01\% when trained on 45\% of the data. However, as the training data ratio decreased, PonziGuard's performance declined significantly, with its BAC dropping to 80.59\% when trained on just 20\% of the data. In contrast, \sys maintained strong performance without any labeled data, demonstrating its robustness and effectiveness. SourceP also showed strong performance but required substantial training data, with a BAC of 92.54\% at 45\% training data, decreasing to 89.39\% at 20\%. SADPonzi, which performed the weakest, exhibited a BAC of 85.59\% with a 45\% training ratio, indicating its difficulty in accurately detecting Ponzi schemes. These results highlight \sys's superior ability to deliver effective detection without extensive training data, maintaining high accuracy and generalizability where other methods require significant labeled data to achieve similar results.

\begin{table}[!h]
	\centering
	\small
	\caption{Performance comparison with \sys (GPT-3.5-turbo), SADPonzi, PonziGuard, and SourceP under different ratio of training data}
	\begin{tabular}{lccccc}
		\hline
		\textbf{Method (Ratio)}    & \textbf{TPR} & \textbf{TNR} & \textbf{FNR} & \textbf{FPR} & \textbf{BAC} \\
		\hline
		\textbf{\sys (0\%)}        & 96.40\%      & 95.71\%      & 3.60\%       & 4.29\%       & 96.06\%      \\
		\textbf{SADPonzi (45\%)}   & 71.94\%      & 99.24\%      & 28.06\%      & 0.76\%       & 85.59\%      \\
		\textbf{PonziGuard (45\%)} & 96.40\%      & 99.62\%      & 3.60\%       & 0.38\%       & 98.01\%      \\
		\textbf{PonziGuard (30\%)} & 73.63\%      & 95.65\%      & 26.37\%      & 4.35\%       & 84.64\%      \\
		\textbf{PonziGuard (20\%)} & 69.07\%      & 92.12\%      & 30.93\%      & 7.88\%       & 80.59\%      \\
		\textbf{SourceP (45\%)}    & 86.67\%      & 98.40\%      & 13.33\%      & 1.60\%       & 92.54\%      \\
		\textbf{SourceP (30\%)}    & 84.91\%      & 98.25\%      & 15.09\%      & 1.75\%       & 91.58\%      \\
		\textbf{SourceP (20\%)}    & 81.05\%      & 97.71\%      & 18.95\%      & 2.29\%       & 89.39\%      \\
		\hline
	\end{tabular}
	\label{tab:comp}
\end{table}

\subsection{RQ3: Ablation Study}
To evaluate the impact of taint analysis and code slicing on detection performance, we conducted an ablation study with two settings. First, we removed the taint propagation graph from the prompts, keeping only the code slices. Second, we removed both taint analysis and slicing, feeding the raw contract code into the model. We then compared the model performances under these conditions.

Table~\ref{tab:aba2} presents the performance of various models without the taint propagation graph. All LLMs showed a decline in detection accuracy. GPT-3.5-turbo experienced a slight decrease in BAC from 96.06\% to 94.27\%.
In contrast, LLAMA2's performance dropped dramatically, with a BAC of 57.82\%, due to its tendency to classify almost all contracts as Ponzi schemes, indicated by a TNR of 15.63\%.
LLAMA3 and Mistral also showed reduced performance, with LLAMA3 achieving a BAC of 78.92\% and Mistral a BAC of 86.41\%, both suffering from increased false negative rates and moderate false positive rates. This suggests that without the taint propagation graph, the models struggle to accurately trace fund flows and identify Ponzi characteristics, leading to higher error rates and demonstrating the importance of taint analysis in enhancing detection accuracy.
\begin{table}[!h]
	\centering
	\small
	\caption{Performance of \sys without taint analysis}
	\begin{tabular}{cccccc}
		\hline
		\textbf{Model}         & \textbf{TPR} & \textbf{TNR} & \textbf{FNR} & \textbf{FPR} & \textbf{BAC} \\
		\hline
		\textbf{GPT-3.5-turbo} & 94.24\%      & 94.29\%      & 5.76\%       & 5.71\%       & 94.27\%      \\
		\textbf{LLAMA2}        & 100\%        & 15.63\%      & 0\%          & 84.37\%      & 57.82\%      \\
		\textbf{LLAMA3}        & 97.12\%      & 60.71\%      & 2.88\%       & 39.29\%      & 78.92\%      \\
		\textbf{Mistral}       & 93.53\%      & 79.29\%      & 6.47\%       & 20.71\%      & 86.41\%      \\
		\hline
	\end{tabular}
	\label{tab:aba2}
\end{table}

Table \ref{tab:aba3} presents the performance of \sys without taint analysis and code slicing. The results show that LLAMA2 continued to exhibit a very high false positive rate, resulting in a low TNR of 13.49\% and a BAC of 56.75\%. Similarly, LLAMA3's false positive rate also increased, leading to a BAC of 78.61\%. In contrast, GPT-3.5-turbo and Mistral showed a decrease in false positive rates but an increase in false negative rates, resulting in BACs of 93.54\% and 90.69\%, respectively. This suggests that without code slicing, the increased input code length affected the models' ability to focus on relevant sections, leading to a higher false negative rate. However, the presence of more context in the input code also helped reduce false positives to some extent. Overall, this ablation study highlights the significant contribution of taint analysis and code slicing in improving detection accuracy, particularly for models like LLAMA2 and LLAMA3, which struggle with longer and more complex inputs.
\begin{table}[!h]
	\centering
	\small
	\caption{Performance of \sys without taint analysis and code slicing}
	\begin{tabular}{cccccc}
		\hline
		\textbf{Model}          & \textbf{TPR} & \textbf{TNR} & \textbf{FNR} & \textbf{FPR} & \textbf{BAC} \\
		\hline
		\textbf{GPT-3.5-turbo } & 90.65\%      & 96.43\%      & 9.35\%       & 3.57\%       & 93.54\%      \\
		\textbf{LLAMA2}         & 100\%        & 13.49\%      & 0\%          & 86.51\%      & 56.75\%      \\
		\textbf{LLAMA3}         & 100\%        & 57.22\%      & 0\%          & 42.78\%      & 78.61\%      \\
		\textbf{Mistral}        & 92.09\%      & 89.29\%      & 7.91\%       & 10.71\%      & 90.69\%      \\
		\hline
	\end{tabular}
	\label{tab:aba3}
\end{table}

\subsection{RQ4: Performance Across Different Versions}

To evaluate the performance of \sys across different Solidity versions and its ability to detect unseen contracts, we used a new, near-balanced dataset comprising 340 newly-collected Ponzi and 300 non-Ponzi contracts, compiled between 2016 and March 2024. This dataset includes many of the latest real-world smart contracts, compiled with Solidity versions ranging from 0.4.11 to 0.8.23. By incorporating a wide range of Solidity versions, this setting helps assess the capability of \sys to generalize and detect Ponzi schemes in contracts that may differ significantly from those seen during training. We employed GPT-3.5-turbo and Mistral as backends for \sys to evaluate their effectiveness in identifying Ponzi contracts under these conditions.

Table~\ref{tab:real_comp} reports the performance of different methods on this dataset. \sys, utilizing GPT-3.5-turbo, achieved the highest balanced accuracy (BAC) of 92.20\%, with a TPR of 92.06\% and a TNR of 92.33\%.PonziGuard is with a BAC of 89.88\%. Notably, when the training data ratio was reduced to 20\%, PonziGuard's performance dropped significantly to a BAC of 77.76\%, illustrating its dependence on larger amounts of training data. In contrast, \sys maintained strong performance without requiring any labeled training data, highlighting its superior ability to detect unseen contracts. Mistral, however, achieved only 50.88\% TPR, significantly lower than GPT-3.5-turbo, due to higher false negatives. This discrepancy is likely due to the longer and more complex contracts in the new dataset, which challenge smaller models like Mistral. Additionally, SourceP performed well with a BAC of 90.98\% but also demonstrated a decline when the training ratio was reduced, emphasizing the importance of training data for its effectiveness. SADPonzi showed the weakest performance with a BAC of 61.78\%, indicating its poor generalization to new contract versions. These results underscore \sys's robustness and superiority in handling complex and unseen contracts across different Solidity versions, particularly when using GPT-3.5-turbo, and demonstrate its effectiveness in detecting diverse Ponzi schemes without extensive training data.

\begin{table}[!h]
	\centering
	\small
	\caption{Performance  comparison with Ours-G (GPT-3.5-turbo), Ours-M (Mistral), SADPonzi, PonziGuard, and SourceP  under different Solidity versions}
	\begin{tabular}{cccccc}
		\hline
		\textbf{Method}                     & \textbf{TPR} & \textbf{TNR} & \textbf{FNR} & \textbf{FPR} & \textbf{BAC} \\
		\hline
		\textbf{Ours-G (0\%)}               & 92.06\%      & 92.33\%      & 7.94\%       & 7.67\%       & 92.20\%      \\
		\textbf{Ours-M (0\%)}               & 50.88\%      & 96.33\%      & 49.12\%      & 3.67\%       & 73.61\%      \\
		\textbf{SADPonzi}            (45\%) & 30.88\%      & 92.67\%      & 69.12\%      & 47.33\%      & 61.78\%      \\
		\textbf{PonziGuard}   (45\%)        & 92.53\%      & 87.23\%      & 7.47\%       & 12.77\%      & 89.88\%      \\
		\textbf{PonziGuard} (30\%)          & 72.55\%      & 86.39\%      & 27.45\%      & 13.61\%      & 79.47\%      \\
		\textbf{PonziGuard}    (20\%)       & 61.76\%      & 93.75\%      & 38.24\%      & 6.25\%       & 77.76\%      \\
		\textbf{SourceP}           (45\%)   & 97.48\%      & 84.47\%      & 2.52\%       & 15.53\%      & 90.98\%      \\
		\textbf{SourceP}          (30\%)    & 67.06\%      & 72.75\%      & 32.94\%      & 27.25\%      & 69.90\%      \\
		\textbf{SourceP}             (20\%) & 33.52\%      & 57.03\%      & 66.48\%      & 42.97\%      & 45.28\%      \\
		\hline
	\end{tabular}
	\label{tab:real_comp}
\end{table}

\subsection{RQ5: Real-world Performance}
To evaluate the effectiveness of \sys in real-world scenarios, we implemented a prototype using GPT-3.5-turbo, and analyzed 4,597 smart contracts verified by Etherscan from March 14-24, 2024. In total, \sys detected 15 new Ponzi schemes from 4,597 contracts, with a total detection cost of \$6.1. These detected Ponzi contracts involved approximately \$20,000 in scam funds. Table~\ref{tab:detected_contracts} provides detailed information about each detected contract's type, amount, deployment date, and transaction details.
We manually verified all detected negative contracts and confirmed all were true negatives. Among detected positives, 28 were false positives, resulting in a FPR of 0.29\%. Of these, 25 were token issuance contracts, and the rest 3 were honeypot contracts, which are fraudulent but not Ponzi schemes.
Although detecting honeypot contracts as Ponzi schemes is a false positive, it reveals the model's potential in identifying various types of fraudulent contracts. This insight underscores the model's capability to reason about the business logic implemented in the source code, identifying contracts that are likely to result in financial loss for investors.

\begin{table}[!h]
	\centering
	\scriptsize
	\caption{Detected Ponzi and honeypot contracts by \sys in March 14-24, 2024 among analyzed 5000 contracts. (T: type, P: Ponzi, H: honeypot.)}
	\begin{tabular}{lcccc}
		\hline
		\textbf{Address}                           & \textbf{T} & ETH & \textbf{Deploy Date} & \textbf{\# Txn} \\
		\hline
		0xa8b9e7718c73329AFd7B99F089C853a80B8127Be & P          & 4   & 2024.03.14           & 6               \\
		0xe713cCf85c89dDc4205747Ed20af7c916094b4Fb & P          & 1.2 & 2024.03.14           & 38              \\
		0x548fD9E32A961404E55Ef25ae08946D9b60fdAD8 & P          & 0.5 & 2024.03.15           & 4               \\
		0x9f40cf150192336023542eAFb7BE723b5B2A306C & P          & 0   & 2024.03.15           & 1               \\
		0xBE5a058d1dB68ffba2c553082A2fd80aAaf0E5dE & P          & 0   & 2024.03.15           & 1               \\
		0xe80Db01D89adE8d035b2aC4507e4899c31145470 & P          & 0   & 2024.03.15           & 1               \\
		0xac1eb84D4b84b13Dde943c89ba7F9CF9343B6778 & P          & 0   & 2024.03.15           & 1               \\
		0xA5aa0241A4E777c2Ba1d98cbd2E1ABa9d5580Cd3 & P          & 0   & 2024.03.15           & 1               \\
		0x35F24D235554912Cb46CdB521b0CbA02c1a61558 & P          & 0   & 2024.03.19           & 1               \\
		0x1bea112a4bA183fcDb3D9B8bc7A144Cb8a01a532 & P          & 0   & 2024.03.19           & 1               \\
		0x1D886858328F7b8504Dd3A90894160bac57EcF1c & P          & 0   & 2024.03.19           & 1               \\
		0x4410fE43B4066Ef89BD76dAd7211C9AA09F155Ee & P          & 0   & 2024.03.19           & 1               \\
		0xa625e957333339b0cceD6Cf5283bF6F5cc988acf & P          & 0   & 2024.03.19           & 1               \\
		0xe94b7f5b6E1deD277Dcaf745A879D2faE8a98fA9 & P          & 0   & 2024.03.19           & 1               \\
		0xEBD042AFc2Bb3B40ECCEc806EeA46cA03afa5635 & P          & 0   & 2024.03.19           & 1               \\
		0xC035BE950a67B2B0e79Bb25dFA826a5970df0D35 & H          & 40  & 2024.03.19           & 3               \\
		0x2748cBCa4fcF6202A8851913f35463a4C75Ef0e9 & H          & 40  & 2024.03.10           & 4               \\
		0x31F625ac650FDa3767aC926B1CD8Dbde569045f4 & H          & 1   & 2024.03.24           & 1               \\
		\hline
	\end{tabular}
	\label{tab:detected_contracts}
\end{table}

We also compared \sys with existing tools in real-world evaluation from March 14 to March 24, 2024. Due to the time-consuming nature of testing these 5000 samples, particularly for tools like SADPonzi and PonziGuard, we randomly sampled \textbf{257} contracts for fair comparing, consisting previous $15$ detected negatives and 242 positives. Table~\ref{tab:detecting_unknown} presents results of different methods in detecting latest real-world contracts.
\sys achieved a perfect TPR of 100\% and a near-perfect TNR of 99.71\%, with FPR of 0\% and a very low FNR of 0.29\%, resulting in a BAC of 99.85\% under the large dataset of 4,597 contracts.
It achieved a BAC of 100\% under the randomly sampled 261 contracts.
In contrast, SADPonzi showed a low TPR of 13.33\% and a TNR of 83.47\%, with a high FPR of 16.53\% and an FNR of 86.67\%, indicating limited accuracy.
PonziGuard performed better with a TPR of 77.78\% and a TNR of 93.42\%, but still exhibited a notable FPR of 6.58\% and an FNR of 22.22\%.
SourceP had a TPR of 66.67\% and a high TNR of 99.07\%, but a significant FPR of 0.93\% and a low FNR of 33.33\%.
Overall, in real-world testing, \sys demonstrated robustness and reliability in detecting previously unknown Ponzi contracts, performing effectively compared to state-of-the-art methods.
\begin{table}[!h]
	\centering
	\small
	\caption{Performance of different methods in detecting real-world newly-deployed contracts (March 14-24, 2024)}
	\begin{tabular}{cccccc}
		\hline
		\textbf{Method}            & \textbf{TPR} & \textbf{TNR} & \textbf{FNR} & \textbf{FPR} & \textbf{BAC} \\
		\hline
		\textbf{SADPonzi(\#257)}   & 13.33\%      & 86.67\%      & 86.67\%      & 16.53\%      & 48.40\%      \\
		\textbf{PonziGuard(\#257)} & 77.78\%      & 93.42\%      & 22.22\%      & 6.58\%       & 85.60\%      \\
		\textbf{SourceP(\#257)}    & 66.67\%      & 99.07\%      & 33.33\%      & 0.93\%       & 82.87\%      \\
		\textbf{\sys(\#257)}       & 100\%        & 100\%        & 0\%          & 0\%          & 100\%        \\
		\textbf{\sys(\#4,597)}     & 100\%        & 99.71\%      & 0\%          & 0.29\%       & 99.85\%      \\
		\hline
	\end{tabular}
	\label{tab:detecting_unknown}
\end{table}

\begin{figure*}[!t]
	\centering
	\subfloat[GPT-3.5-turbo time]{\includegraphics[width = 0.17\linewidth]{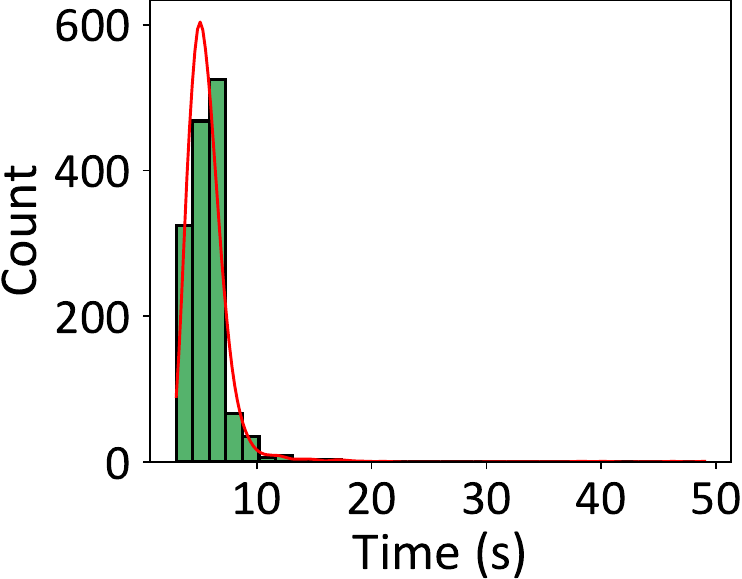}}
	\subfloat[LLAMA2 time]{\includegraphics[width = 0.17\linewidth]{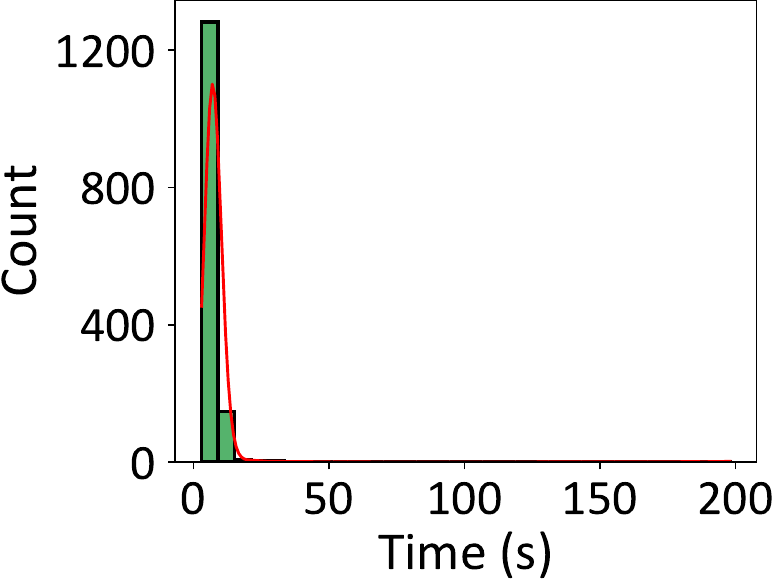}}
	\subfloat[LLAMA3 time]{\includegraphics[width = 0.17\linewidth]{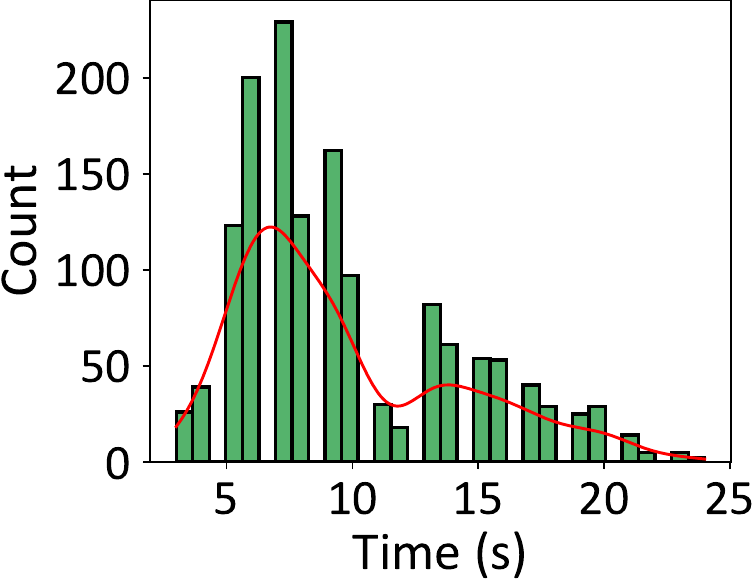}}
	\subfloat[Mistral time]{\includegraphics[width = 0.168\linewidth]{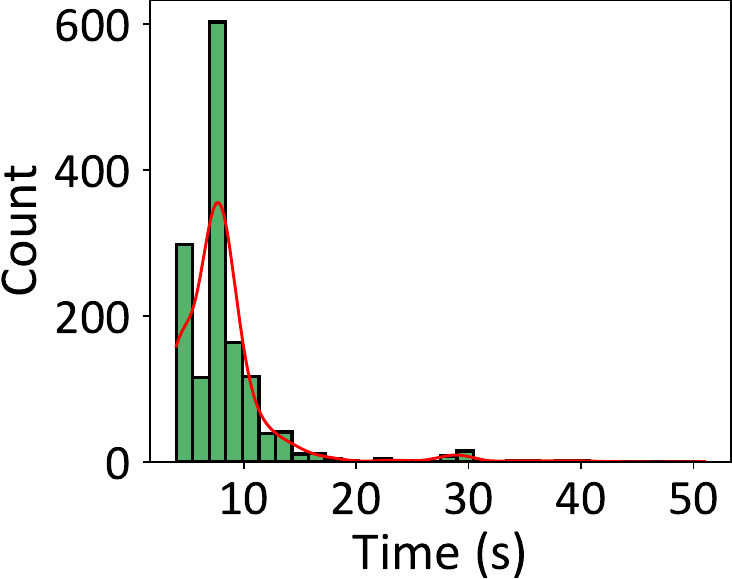}}
	\subfloat[GPT-3.5-turbo cost]{\includegraphics[width = 0.17\linewidth]{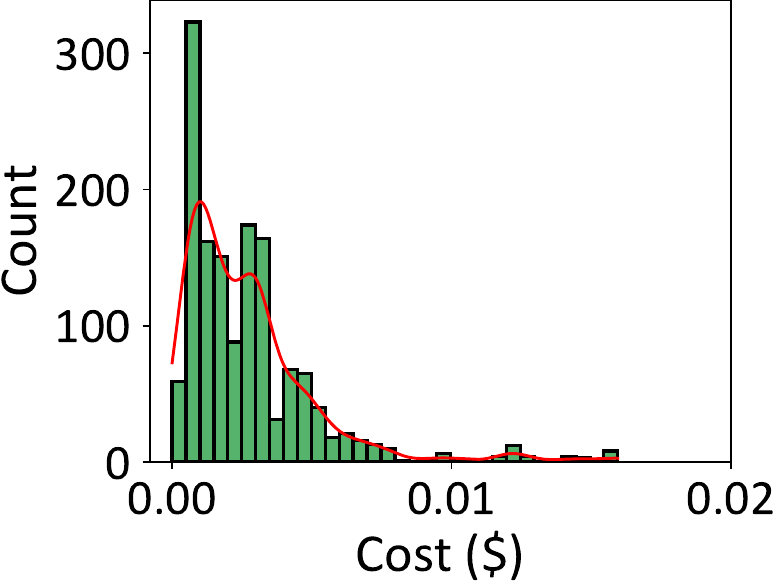}}
	\subfloat[GPT-3.5-turbo token]{\includegraphics[width = 0.17\linewidth]{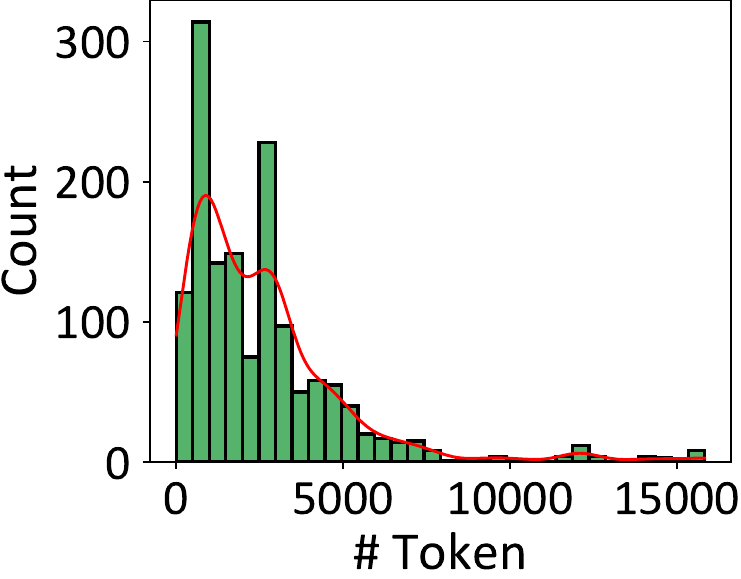}}
	\caption{Distribution of overhead under GPT-3.5-turbo (a,e,f), LLAMA2 (b), LLAMA3 (c), Mistral (d)}
	\label{fig:overheads}
\end{figure*}

\subsection{RQ6. Overhead Estimation}
We also evaluated \sys's overhead in different settings.
Table~\ref{tab:overhead1} reports the overall overhead of \sys, demonstrating that GPT-3.5-turbo is highly efficient, with an average processing time of 5.52 seconds per contract, using 2601.3 tokens per detection at a cost of \$0.0027 per contract. This efficiency is noteworthy given the complexity involved in the two-step prompt process. In comparison, LLAMA2 and LLAMA3 models require significantly more time, averaging 7.85 seconds and 9.68 seconds per contract, respectively, though specific token usage and cost metrics were not applicable (n.a) for these models due to their open-source nature. Mistral, another open-source model, performs similarly to GPT-3.5-turbo with an average processing time of 7.71 seconds, indicating its capability to handle complex prompts effectively despite lacking specific token and cost metrics.
Figure~\ref{fig:overheads} presents distribution of overhead under different models in detail.

\begin{table}[!h]
	\centering
	\small
	\caption{Overhead of \sys}
	\begin{tabular}{cccc}
		\hline
		\textbf{Model}         & \textbf{Mean/std time(s)} & \textbf{Mean tokens} & \textbf{Mean cost(\$)} \\
		\hline
		\textbf{GPT-3.5-turbo} & 5.52/8.64                 & 2601.3               & 0.0027                 \\
		\textbf{LLAMA2}        & 7.85/127.16               & n.a                  & n.a                    \\
		\textbf{LLAMA3}        & 9.68/20.01                & n.a                  & n.a                    \\
		\textbf{Mistral}       & 7.71/29.97                & n.a                  & n.a                    \\
		\hline
	\end{tabular}
	\label{tab:overhead1}
\end{table}

Table \ref{tab:overhead2} reports overhead of \sys without taint analysis.
GPT-3.5-turbo demonstrates excellent efficiency with an average time of 1.52 seconds per contract, utilizing 1753.6 tokens at a cost of \$0.0012 per detection. This represents a significant reduction in processing time compared to the complete overhead configuration. LLAMA2 also shows improved performance, with an average time of 1.21 seconds, indicating better efficiency when handling simpler prompts. In contrast, LLAMA3 remains relatively slow, averaging 5.88 seconds per contract, which suggests inefficiencies even in the absence of taint analysis. Mistral performs well, with an average time of 1.61 seconds, close to GPT-3.5-turbo, indicating its effectiveness in handling one-step prompts efficiently.

\begin{table}[!h]
	\centering
	\small
	\caption{Overhead of \sys without taint analysis}
	\begin{tabular}{cccc}
		\hline
		\textbf{Model}         & \textbf{Mean/std time(s)} & \textbf{Mean tokens} & \textbf{Mean cost(\$)} \\
		\hline
		\textbf{GPT-3.5-turbo} & 1.52/0.87                 & 1753.6               & 0.0012                 \\
		\textbf{LLAMA2}        & 1.21                      & n.a                  & n.a                    \\
		\textbf{LLAMA3}        & 5.88/22.42                & n.a                  & n.a                    \\
		\textbf{Mistral}       & 1.61/1.16                 & n.a                  & n.a                    \\
		\hline
	\end{tabular}
	\label{tab:overhead2}
\end{table}

Table~\ref{tab:overhead3} illustrates the performance under the raw code input configuration. GPT-3.5-turbo exhibits the fastest average time of 1.08 seconds per contract, but it uses 3,000 tokens per detection, increasing the cost to \$0.003 per contract. This suggests a trade-off between speed and cost when dealing with raw code. LLAMA2 shows the quickest performance with an average time of 1.28 seconds, highlighting its efficiency with raw code, though token and cost data are not available. LLAMA3 again shows higher time consumption with an average of 12.01 seconds, confirming its slower processing capabilities. Mistral performs efficiently with an average time of 1.07 seconds, similar to GPT-3.5-turbo, indicating its capability to handle raw code effectively.
\begin{table}[!h]
	\centering
	\small
	\caption{Overhead of \sys without taint analysis and code slicing}
	\begin{tabular}{cccc}
		\hline
		\textbf{Model}         & \textbf{Mean/std time(s)} & \textbf{Mean tokens} & \textbf{Mean cost(\$)} \\
		\hline
		\textbf{GPT-3.5-turbo} & 1.08/0.12                 & 3470.5               & 0.0033                 \\
		\textbf{LLAMA2}        & 1.28/0.22                 & n.a                  & n.a                    \\
		\textbf{LLAMA3}        & 12.01/6.48                & n.a                  & n.a                    \\
		\textbf{Mistral}       & 1.07/0.12                 & n.a                  & n.a                    \\
		\hline
	\end{tabular}
	\label{tab:overhead3}
\end{table}

\subsection{Real Case Study}
We also presented three case study of detected Ponzi contracts.

\textbf{Contract 0xa8..7Be}~\cite{0xa8b9e7718c} has three main functions:
\texttt{constructor},
\texttt{receive}, and
\texttt{processPayOut}, as depicted in Listing~\ref{code:0xa8}.
\texttt{constructor} sets the creator as the owner and initializes the Ray structure with the owner's address and specific amounts. \texttt{receive} function processes user investments, assigns investment levels based on the amount, updates relevant indices and funds, and records the investor's details. \texttt{processPayOut} function distributes returns to investors within a specified level until all are paid or the funds are exhausted, ensuring proper updates to indices and remaining funds. This contract aims to manage and automate the distribution of investment returns systematically.

\begin{lstlisting}[caption={Code snippet of contract 0xa8..7Be},label={code:0xa8} ] 
	constructor() {...
		owner = msg.sender;ray[0][0]=Ray(msg.sender,8710000 gwei); 
		ray[1][0] = Ray(msg.sender, 18781000 gwei);...
		ray[9][0] = Ray(msg.sender, 871780000000000 gwei); }
	function receive() external payable nonReentrant {
		if (msg.value > 871000 gwei) {...
			uint256 value = msg.value + GAS_PER_ITERATION * gasPrice;
			if (value <= 8710000 gwei) {level = 0;} else if (value <= 87100000 gwei) { level = 1;...} else {
				address payable receiver = payable(msg.sender);
				Address.sendValue(receiver, msg.value);	return;	}
			if (msg.value < GAS_MIN_UTILIZE * gasPrice) {...
				range[level].lastIndex++; value=msg.value+gasCost;
				range[level].currentFunds+=msg.value; 
				sumFunds+=msg.value;
				ray[level][range[level].lastIndex]=Ray({addr: msg.sender, utilize: value}); 
				emit UtilizeReceived(msg.sender,value,level,range[level].lastIndex);}...}}
	function processPayOut(value,sumF,leftCur,level,counterIt,maxCounter) {
		while (counterIt < maxCounter && value >= leftCur && range[level].firstIndex <= range[level].lastIndex) {...
			address payable receivePayout = payable(ray[level][range[level].firstIndex].addr);
			Address.sendValue(receivePayout, leftCur);
			emit UtilizePayOut(ray[level][range[level].firstIndex].addr,leftCur,level,range[level].firstIndex);
			delete ray[level][range[level].firstIndex];
			range[level].firstIndex++; counterIt++;...}...
		range[level].currentFunds = value; range[level].leftFunds = leftCur;
		sumFunds = sumF; return counterIt;}...}
		\end{lstlisting}

		\textbf{Contract 0xe7..4Fb}~\cite{0xe713cCf8} exhibits characteristics of a Ponzi scheme, particularly through its \texttt{buyTokens} and \texttt{payReferRew} functions, as depicted in Listing~\ref{code:0xe7}. These functions incentivize the recruitment of new investors by offering referral rewards, creating a pyramid-like structure where referrers earn commissions up to three levels deep. \texttt{setCurrentPhase} function manipulates the price of the Sale token to appear more valuable as more tokens are sold, enticing further investment. Additionally, \texttt{withdrawFunds} allows only the contract creator to withdraw all remaining funds, a common trait in Ponzi schemes to facilitate an exit scam. Unlike traditional Ponzi contracts using Ether, this contract uses USDT, promising returns in USDT tokens, thus disguising its true nature and making it harder to identify as a Ponzi scheme. This mechanism attracts investors under the guise of token trading, but the underlying referral and fund manipulation tactics reveal its fraudulent intent.
		
		\begin{lstlisting}[caption={Code snippet of contract 0xe7..4Fb},,label={code:0xe7} ]
function buyTokens(amount, _referralAddress) {...
	USDT.safeTransferFrom(msg.sender, address(this), remainingUSD / scalingDivisor);
	if (referralAddress[msg.sender]==address(0)) {
		referralAddress[msg.sender]=_referralAddress; 
		myReferrals[_referralAddress].push(msg.sender);}
	_referralAddress=referralAddress[msg.sender];
	_totalUSDTInvestment+=remainingUSD;
	setCurrentPhase(tokenSold);
	uint256 _remainingTokens=payReferRew(_referralAddress, remainingUSD);	...	}
function withdrawFunds() public onlyOwner{...
	USDT.safeTransfer(Funds_Wallet, tokenCollected );
	tokenCollected=0;}
function setCurrentPhase(_tokens) {...
	for(i=0;i<phaseLimit.length;i++) {if(_tokens>phaseLimit[i]){currentPhase=i+1;}}}
function payReferRew(_referralAddress, _amount) {...
	for(i=0;i<referLevels.length&&uplineAddress!=address(0);i++) {
		uint256 _rewards=(_amount * referLevels[i])/100; 
		referralRewards[uplineAddress]+=_rewards;	rewPaid += _rewards;
		setReferredUser(i + 1, uplineAddress);...}
	return (_amount - rewPaid);}...}
\end{lstlisting}

\begin{lstlisting}[caption={Code snippet of contract 0x96..f27},,label={code:0x96} ]
function transferFrom(sender,recipient,amount){...
uint256 currentAllowance=_allowances[sender][_msgSender()];
_transfer(sender,recipient, amount);
_approve(sender,_msgSender(), currentAllowance-amount);
return true;}
function sendETHToINTEL(amount) {...
taxOperator.transfer(amount/2); 
teamOperator.transfer(amount/2);}
function recoverERC20(_tokenAddy,_amount) onlyOwner {...
require(_amount<=IERC20(_tokenAddy).balanceOf(address(this)),"Insufficient Amount");
IERC20(_tokenAddy).transfer(teamOperator, _amount);
emit ERC20TokensRecovered(_amount);}
function recoverETH() external {...
uint256 contractETHs=address(this).balance;
payable(address(teamOperator)).transfer(contractETHs);
\colorbox{highlight}{emit ETHBalanceRecovered()};}
\end{lstlisting}

\begin{figure}[!t]
\centering
\includegraphics[width = 0.7\linewidth]{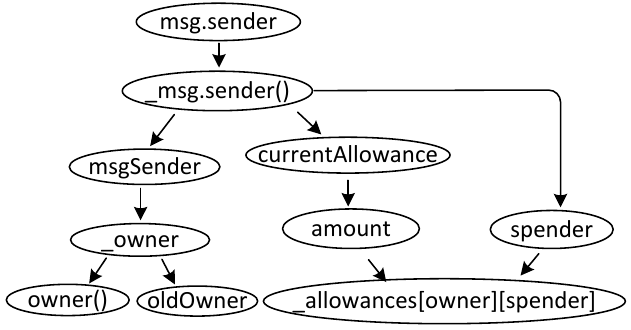}
\caption{Taint propagation graph of contract 0x96..f27}
\label{fig:taintp}
\end{figure}

\textbf{Contract 0x96..f27}~\cite{0x96Da8b9cfEC} is a false positive case identified by \sys, with its code shown in Listing~\ref{code:0x96} and its taint propagation graph depicted in Figure~\ref{fig:taintp}.
It has functions that resemble Ponzi scheme operations, such as distributing ETH and recovering tokens, but these actions are not funded by new investor money, a key Ponzi characteristic. The \texttt{sendETHToINTEL} function distributes ETH in a predefined manner, not reliant on new investments. Similarly, the \texttt{recoverERC20} and \texttt{recoverETH} functions allow the owner to recover tokens and ETH, which could be misinterpreted as fund funneling, but they do not inherently indicate a Ponzi scheme without the context of paying returns with new investor funds. The presence of tax fees and trading requirements may suggest reliance on new investors, but these features can also be legitimate. The taint analysis failed to identify Ether reallocation consistent with Ponzi schemes, resulting in a false positive. The detected results by \sys is given as follows:

\noindent \fbox{
\begin{minipage}{0.98\linewidth}\small
Based on the provided smart contract code, there are several concerning elements that align with the features and patterns of a Ponzi scheme:
1. The \texttt{sendETHToINTEL} function appears to \textcolor{blue}{\textbf{distribute ETH to specific addresses}}, which could be indicative of unsustainable payout structures.
2. The \texttt{recoverERC20} and \texttt{recoverETH} functions allow the \textcolor{blue}{\textbf{owner to recover tokens and ETH from the contract}}, potentially indicating a \textcolor{blue}{\textbf{dependency on new investments to sustain payouts}}.
3. The presence of \textcolor{blue}{\textbf{buy and sell tax fees}}, as well as the \textcolor{blue}{\textbf{requirement for trading to be enabled}}, could suggest a \textcolor{blue}{\textbf{ reliance on new investors to generate returns for existing participants}}.
4. The \textcolor{blue}{\textbf{lack of clear revenue-generating activities}} or legitimate investment strategies raises concerns about the \textcolor{blue}{\textbf{source of potential returns for investors}}.
Based on the analysis of the provided smart contract code, there are indications that it exhibits features and patterns consistent with a Ponzi scheme. It is important to conduct a thorough review of the entire smart contract and its associated documentation to fully assess its nature and potential risks.
\end{minipage}}

\section{Related Work}
\textbf{Ponzi schemes analysis.}
Ponzi schemes, classic financial frauds, have evolved with technological advancements, leading to the proliferation of online high-yield investment programs and a surge of blockchain-based Ponzi schemes.
Early research by Chen et al.\cite{chen2018detecting} and Fan et al.\cite{fan2021spsd} introduced machine learning methods, such as extracting features from user accounts, contract opcodes, and using CatBoost, to identify Ponzi schemes in Ethereum smart contracts.
Notable contributions include SADPonzi~\cite{chen2021sadponzi} identifying Ponzi schemes via bytecode, and MulCas~\cite{zheng2023securing}, a multi-view cascade detection model. Liang et al.\cite{liang2024ponziguard,liang2024towards} employed contract runtime behavior graphs.
Efforts have also been made to detect Ponzi schemes on Bitcoin and other blockchain platforms, highlighting their widespread nature. Smart contract analysis has extensively focused on identifying security vulnerabilities in Ethereum, with tools like Oyente\cite{luu2016making}, Osiris~\cite{torres2018osiris}, Mythril~\cite{mueller2018smashing}, Maian~\cite{nikolic2018finding}, and Manticore~\cite{mossberg2019manticore} using symbolic execution, while Securify and Zeus use formal verification methods.
Tools such as Slither~\cite{feist2019slither} and SmartCheck~\cite{tikhomirov2018smartcheck} utilize static analysis, and ContractFuzzer~\cite{jiang2018contractfuzzer} and ReGuard~\cite{liu2018reguard} employ fuzz testing for vulnerability detection.

Comprehensive empirical reviews of these tools have been conducted by researchers like Thomas et al.~\cite{durieux2020empirical}, while Pinna et al.~\cite{pinna2019massive} provided an empirical study  of Ethereum smart contracts. Additional studies have explored code smells, gas usage, and code cloning behaviors in smart contracts. Bartoletti et al.~\cite{bartoletti2020dissecting} were among the first to study Ponzi schemes on Ethereum, using Normalized Levenshtein Distance (NLD) to measure contract bytecode similarity. Rule-based approaches by Sun et al.~\cite{sun2020early} and Chen et al.~\cite{chen2021sadponzi} leverage behavior forest similarity and symbolic execution, respectively. However, these methods rely heavily on expert knowledge and predefined rules, limiting their ability to detect unknown Ponzi schemes. Static information-based approaches~\cite{yu2021ponzi,lou2020ponzi}, such as opcode frequency and transaction data, also face challenges in distinguishing Ponzi contracts due to their reliance on static data, which fails to capture the dynamic nature of such schemes.

\textbf{Smart contract fuzzing and taint analysis.}
Fuzzing has been effective in uncovering vulnerabilities in smart contracts. ContractFuzzer~\cite{jiang2018contractfuzzer} is a black-box fuzzer for detecting security bugs in Ethereum contracts. Grey-box fuzzers have also been developed to exploit vulnerabilities. While these methods are primarily aimed at identifying security flaws, PonziGuard~\cite{liang2024ponziguard} utilizes fuzzing to invoke contracts and gather runtime information for detecting Ponzi schemes.
Taint analysis is a powerful technique for examining data flow in programs. It has been used in smart contract analysis by tools like Osiris~\cite{torres2018osiris}, Sereum~\cite{rodler2018sereum}, and EthPloit~\cite{zhang2020ethploit}.
Osiris~\cite{torres2018osiris} combines taint analysis with symbolic execution to detect integer bugs.
Sereum~\cite{rodler2018sereum} uses taint analysis to protect against reentrancy vulnerabilities,
and EthPloit~\cite{zhang2020ethploit} generates exploit-targeted transaction sequences to enhance contract fuzzing efficiency. These studies focus on security vulnerabilities, while our tool specifically targets the identification of malicious contracts, including Ponzi schemes.

\textbf{LLM-based code analysis.}
Code analysis is crucial for identifying vulnerabilities and ensuring code quality, and recent advancements in natural language processing, particularly transformers and LLMs, have significantly enhanced the detection of malicious activities like Ponzi schemes in smart contracts. For instance, Chen et al.~\cite{chiang2023can} utilized transformer-based methods to improve the readability of decompiled code by predicting variable names and types, while Xu et al.~\cite{xu2023lmpa} showed that iterative algorithms with multiple LLM queries could enhance decompilation results. Comprehensive reviews by Bubeck et al.~\cite{bubeck2023sparks} and Chen et al.~\cite{chen2021evaluating} have demonstrated the potential of LLMs in coding and security, particularly in code generation and analysis. LLMs excel in detecting Ponzi schemes in smart contracts due to their ability to understand complex code structures. For instance, GPTScan~\cite{sun2024gptscan} combines GPT with static analysis to detect logic vulnerabilities in smart contracts, achieving high precision and recall rates. By matching candidate vulnerabilities with GPT and validating them through static confirmation, GPTScan effectively reduces false positives. This highlights the potential of LLMs to enhance smart contract security and detect complex vulnerabilities, making them a powerful tool against blockchain-based Ponzi schemes.

\section{Discussion}
\textbf{Consistent LLM performance across multiple-runs.}
In our evaluation, we ran each prompt five times and computed the mean performance to account for potential variability in responses, as LLMs are known to produce different outputs even with a fixed temperature setting. Interestingly, we observed that the LLM's performance was more consistent than anticipated, with minimal deviation across the runs. This consistency suggests that, while variability is a known issue, the impact may be less significant in structured tasks like Ponzi detection. Therefore, our results reflect a reliable assessment of the model's capabilities in this context.

\textbf{Comparing with existing work.}
We excluded certain related works, such as those relying on analysis methods~\cite{chen2018detecting,fan2021spsd} or transaction patterns~\cite{cai2023ponzi,yu2021ponzi}, from our experimental comparison due to their performance and accessibility issues. These works did not achieve performance levels comparable to the state-of-the-art methods we evaluated. Additionally, the lack of open-source code for many of these approaches made it difficult to replicate their results accurately. Despite our attempts to reproduce their claimed performance using the available information, we were unsuccessful. Therefore, we chose not to include these works in our comparison to ensure the accuracy and reliability of our evaluation.

\textbf{Necessarity of LLM in decision step.}
An LLM is essential in the final step  because it excels at understanding and interpreting the intricate, often ambiguous nature of smart contracts, which can involve sophisticated and evolving schemes that traditional algorithms might overlook. Unlike rule-based systems that require explicit programming for every possible scenario, an LLM can adapt to new contexts, detecting patterns and behaviors that deviate from known Ponzi schemes but still exhibit fraudulent characteristics. This adaptability is particularly valuable in the dynamic environment of blockchain, where bad actors continuously innovate to bypass conventional detection methods. By leveraging LLM's ability to synthesize complex information and draw on a vast knowledge base, we can achieve a more nuanced and accurate detection of Ponzi schemes, even in cases where algorithmic approaches would fail.

\section{Conclusion}

This work introduces \sys, a novel LLM-driven framework for detecting Ponzi smart contracts with minimal labeled data. By leveraging advanced language models, zero-shot chain-of-thought prompting, static taint analysis, and automated code slicing, \sys effectively uncovers the complex fraud patterns in Ponzi schemes.
Our comprehensive evaluation shows that \sys excels in detecting unseen Ponzi contracts, achieving high accuracy and efficiency in both controlled and real-world settings.
It offers a powerful and adaptable solution for real-time Ponzi contract monitoring, ensuring robust and scalable detection capabilities in the evolving landscape of blockchain security.

\bibliographystyle{abbrv}
\balance
\bibliography{ref.bib}

\begin{thebibliography}{10}

\bibitem{0x2a53f4}
Contract 0x2a53f42ad8bba138c21b50a4e5711f18381a61e9.
\newblock
  \url{https://etherscan.io/address/0x2a53f42ad8bba138c21b50a4e5711f18381a61e9},
  2024.

\bibitem{0x96Da8b9cfEC}
Contract 0x96da8b9cfec99a1ccff16ab16f3948da82396f27.
\newblock
  \url{https://etherscan.io/address/0x96Da8b9cfEC99A1CcFF16AB16F3948dA82396f27#code},
  2024.

\bibitem{0xa8b9e7718c}
Contract 0xa8b9e7718c73329afd7b99f089c853a80b8127be.
\newblock
  \url{https://etherscan.io/address/0xa8b9e7718c73329AFd7B99F089C853a80B8127Be#code},
  2024.

\bibitem{0xe713cCf8}
Contract 0xe713ccf85c89ddc4205747ed20af7c916094b4fb.
\newblock
  \url{https://etherscan.io/address/0xe713cCf85c89dDc4205747Ed20af7c916094b4Fb#code},
  2024.

\bibitem{gpt35}
https://platform.openai.com/docs/models/gpt-3-5-turbo.
\newblock \url{https://platform.openai.com/docs/models/gpt-3-5-turbo}, 2024.

\bibitem{Llama2}
Llama 2.
\newblock \url{https://github.com/meta-llama/llama}, 2024.

\bibitem{Llama3}
Llama 3.
\newblock \url{https://github.com/meta-llama/llama3}, 2024.

\bibitem{Mistral}
Mistral 7b.
\newblock \url{https://github.com/mistralai/mistral-inference}, 2024.

\bibitem{xblock}
Ponzi contract dataset.
\newblock \url{https://xblock.pro/#/dataset/25}, 2024.

\bibitem{PonziDataset}
Ponzidataset.
\newblock \url{https://github.com/smartcontract-detect-yzu/PonziDataset}, 2024.

\bibitem{slither}
Slither, the smart contract static analyzer.
\newblock \url{https://github.com/crytic/slither}, 2024.

\bibitem{solc}
solc-bin.
\newblock \url{https://github.com/ethereum/solc-bin}, 2024.

\bibitem{bartoletti2020dissecting}
M.~Bartoletti, S.~Carta, T.~Cimoli, and R.~Saia.
\newblock Dissecting ponzi schemes on ethereum: identification, analysis, and
  impact.
\newblock {\em Future Generation Computer Systems}, 2020.

\bibitem{bubeck2023sparks}
S.~Bubeck, V.~Chandrasekaran, R.~Eldan, J.~Gehrke, E.~Horvitz, E.~Kamar,
  P.~Lee, Y.~T. Lee, Y.~Li, S.~Lundberg, et~al.
\newblock Sparks of artificial general intelligence: Early experiments with
  gpt-4.
\newblock {\em arXiv:2303.12712}, 2023.

\bibitem{cai2023ponzi}
J.~Cai, B.~Li, J.~Zhang, and X.~Sun.
\newblock Ponzi scheme detection in smart contract via transaction semantic
  representation learning.
\newblock {\em IEEE Transactions on Reliability}, 2023.

\bibitem{chen2021evaluating}
M.~Chen, J.~Tworek, H.~Jun, Q.~Yuan, H.~P. d.~O. Pinto, J.~Kaplan, H.~Edwards,
  Y.~Burda, N.~Joseph, G.~Brockman, et~al.
\newblock Evaluating large language models trained on code.
\newblock {\em arXiv:2107.03374}, 2021.

\bibitem{chen2021sadponzi}
W.~Chen, X.~Li, Y.~Sui, N.~He, H.~Wang, L.~Wu, and X.~Luo.
\newblock Sadponzi: Detecting and characterizing ponzi schemes in ethereum
  smart contracts.
\newblock {\em ACM on Measurement and Analysis of Computing Systems}, 2021.

\bibitem{chen2018detecting}
W.~Chen, Z.~Zheng, J.~Cui, E.~Ngai, P.~Zheng, and Y.~Zhou.
\newblock Detecting ponzi schemes on ethereum: Towards healthier blockchain
  technology.
\newblock In {\em The World Wide Web Conference (WWW)}, 2018.

\bibitem{chen2024ponzifinder}
Y.~Chen, B.~Li, Y.~Xiao, and X.~Du.
\newblock Ponzifinder: Attention-based edge-enhanced ponzi contract detection.
\newblock {\em IEEE Transactions on Reliability}, 2024.

\bibitem{chiang2023can}
C.-H. Chiang and H.-y. Lee.
\newblock Can large language models be an alternative to human evaluations?
\newblock {\em arXiv:2305.01937}, 2023.

\bibitem{durieux2020empirical}
T.~Durieux, J.~F. Ferreira, R.~Abreu, and P.~Cruz.
\newblock Empirical review of automated analysis tools on 47,587 ethereum smart
  contracts.
\newblock In {\em ACM/IEEE International Conference on Software Engineering
  (ICSE)}, 2020.

\bibitem{fan2021spsd}
S.~Fan, S.~Fu, H.~Xu, and X.~Cheng.
\newblock Al-spsd: Anti-leakage smart ponzi schemes detection in blockchain.
\newblock {\em Information Processing and Management}, 2021.

\bibitem{fang2024automated}
Z.~Fang, Z.~Lin, Z.~Chen, X.~Chen, Y.~Gao, and Y.~Fang.
\newblock Automated federated pipeline for parameter-efficient fine-tuning of
  large language models.
\newblock {\em arXiv preprint arXiv:2404.06448}, 2024.

\bibitem{fang2024ic3m}
Z.~Fang, Z.~Lin, S.~Hu, H.~Cao, Y.~Deng, X.~Chen, and Y.~Fang.
\newblock Ic3m: In-car multimodal multi-object monitoring for abnormal status
  of both driver and passengers.
\newblock {\em arXiv preprint arXiv:2410.02592}, 2024.

\bibitem{feist2019slither}
J.~Feist, G.~Grieco, and A.~Groce.
\newblock Slither: a static analysis framework for smart contracts.
\newblock In {\em International Workshop on Emerging Trends in Software
  Engineering for Blockchain}, pages 8--15. IEEE, 2019.

\bibitem{galletta2024explainable}
L.~Galletta and F.~Pinelli.
\newblock Explainable ponzi schemes detection on ethereum.
\newblock In {\em ACM/SIGAPP Symposium on Applied Computing}, 2024.

\bibitem{jiang2018contractfuzzer}
B.~Jiang, Y.~Liu, and W.~K. Chan.
\newblock Contractfuzzer: Fuzzing smart contracts for vulnerability detection.
\newblock In {\em ACM/IEEE International Conference on Automated Software
  Engineering (ASE)}, pages 259--269, 2018.

\bibitem{liang2024ponziguard}
R.~Liang, J.~Chen, K.~He, Y.~Wu, G.~Deng, R.~Du, and C.~Wu.
\newblock Ponziguard: Detecting ponzi schemes on ethereum with contract runtime
  behavior graph (crbg).
\newblock In {\em ACM/IEEE International Conference on Software Engineering
  (ICSE)}, 2024.

\bibitem{liang2024vulseye}
R.~Liang, J.~Chen, C.~Wu, K.~He, Y.~Wu, R.~Cao, R.~Du, Y.~Liu, and Z.~Zhao.
\newblock Vulseye: Detect smart contract vulnerabilities via stateful directed
  graybox fuzzing.
\newblock {\em arXiv preprint arXiv:2408.10116}, 2024.

\bibitem{liang2024towards}
R.~Liang, J.~Chen, C.~Wu, K.~He, Y.~Wu, W.~Sun, R.~Du, Q.~Zhao, and Y.~Liu.
\newblock Towards effective detection of ponzi schemes on ethereum with
  contract runtime behavior graph.
\newblock {\em arXiv preprint arXiv:2406.00921}, 2024.

\bibitem{lin2024fedsn}
Z.~Lin, Z.~Chen, Z.~Fang, X.~Chen, X.~Wang, and Y.~Gao.
\newblock Fedsn: A federated learning framework over heterogeneous leo
  satellite networks.
\newblock {\em IEEE Transactions on Mobile Computing}, 2024.

\bibitem{lin2024splitlora}
Z.~Lin, X.~Hu, Y.~Zhang, Z.~Chen, Z.~Fang, X.~Chen, A.~Li, P.~Vepakomma, and
  Y.~Gao.
\newblock Splitlora: A split parameter-efficient fine-tuning framework for
  large language models.
\newblock {\em arXiv preprint arXiv:2407.00952}, 2024.

\bibitem{lin2023pushing}
Z.~Lin, G.~Qu, Q.~Chen, X.~Chen, Z.~Chen, and K.~Huang.
\newblock Pushing large language models to the 6g edge: Vision, challenges, and
  opportunities.
\newblock {\em arXiv preprint arXiv:2309.16739}, 2023.

\bibitem{lin2024adaptsfl}
Z.~Lin, G.~Qu, W.~Wei, X.~Chen, and K.~K. Leung.
\newblock Adaptsfl: Adaptive split federated learning in resource-constrained
  edge networks.
\newblock {\em arXiv preprint arXiv:2403.13101}, 2024.

\bibitem{lin2024efficient}
Z.~Lin, G.~Zhu, Y.~Deng, X.~Chen, Y.~Gao, K.~Huang, and Y.~Fang.
\newblock Efficient parallel split learning over resource-constrained wireless
  edge networks.
\newblock {\em IEEE Transactions on Mobile Computing}, 2024.

\bibitem{liu2018reguard}
C.~Liu, H.~Liu, Z.~Cao, Z.~Chen, B.~Chen, and B.~Roscoe.
\newblock Reguard: finding reentrancy bugs in smart contracts.
\newblock In {\em ACM/IEEE International Conference on Software Engineering
  (ICSE)}, 2018.

\bibitem{lou2020ponzi}
Y.~Lou, Y.~Zhang, and S.~Chen.
\newblock Ponzi contracts detection based on improved convolutional neural
  network.
\newblock In {\em IEEE International Conference on Services Computing (SCC)},
  2020.

\bibitem{lu2024sourcep}
P.~Lu, L.~Cai, and K.~Yin.
\newblock Sourcep: Detecting ponzi schemes on ethereum with source code.
\newblock In {\em IEEE International Conference on Acoustics, Speech and Signal
  Processing (ICASSP)}, 2024.

\bibitem{luu2016making}
L.~Luu, D.-H. Chu, H.~Olickel, P.~Saxena, and A.~Hobor.
\newblock Making smart contracts smarter.
\newblock In {\em ACM SIGSAC Conference on Computer and Communications Security
  (CCS)}, 2016.

\bibitem{mossberg2019manticore}
M.~Mossberg, F.~Manzano, E.~Hennenfent, A.~Groce, G.~Grieco, J.~Feist,
  T.~Brunson, and A.~Dinaburg.
\newblock Manticore: A user-friendly symbolic execution framework for binaries
  and smart contracts.
\newblock In {\em IEEE/ACM International Conference on Automated Software
  Engineering (ASE)}, 2019.

\bibitem{mueller2018smashing}
B.~Mueller.
\newblock Smashing ethereum smart contracts for fun and real profit.
\newblock {\em HITB SECCONF Amsterdam}, 9:54, 2018.

\bibitem{nikolic2018finding}
I.~Nikoli{\'c}, A.~Kolluri, I.~Sergey, P.~Saxena, and A.~Hobor.
\newblock Finding the greedy, prodigal, and suicidal contracts at scale.
\newblock In {\em Annual Computer Security Applications Conference (ACSAC)},
  pages 653--663, 2018.

\bibitem{pinna2019massive}
A.~Pinna, S.~Ibba, G.~Baralla, R.~Tonelli, and M.~Marchesi.
\newblock A massive analysis of ethereum smart contracts empirical study and
  code metrics.
\newblock {\em Ieee Access}, 2019.

\bibitem{rodler2018sereum}
M.~Rodler, W.~Li, G.~O. Karame, and L.~Davi.
\newblock Sereum: Protecting existing smart contracts against re-entrancy
  attacks.
\newblock In {\em NDSS}, 2018.

\bibitem{sun2020early}
W.~Sun, G.~Xu, Z.~Yang, and Z.~Chen.
\newblock Early detection of smart ponzi scheme contracts based on behavior
  forest similarity.
\newblock In {\em International Conference on Software Quality, Reliability and
  Security (QRS)}, 2020.

\bibitem{sun2024gptscan}
Y.~Sun, D.~Wu, Y.~Xue, H.~Liu, H.~Wang, Z.~Xu, X.~Xie, and Y.~Liu.
\newblock Gptscan: Detecting logic vulnerabilities in smart contracts by
  combining gpt with program analysis.
\newblock In {\em ACM/IEEE International Conference on Software Engineering
  (ICSE)}, pages 1--13, 2024.

\bibitem{sun2023panda}
Z.~Sun, X.~Luo, and Y.~Zhang.
\newblock Panda: Security analysis of algorand smart contracts.
\newblock In {\em USENIX Security Symposium}, 2023.

\bibitem{tikhomirov2018smartcheck}
S.~Tikhomirov, E.~Voskresenskaya, I.~Ivanitskiy, R.~Takhaviev, E.~Marchenko,
  and Y.~Alexandrov.
\newblock Smartcheck: Static analysis of ethereum smart contracts.
\newblock In {\em International Workshop on Emerging Trends in Software
  Engineering for Blockchain}, pages 9--16, 2018.

\bibitem{torres2018osiris}
C.~F. Torres, J.~Sch{\"u}tte, and R.~State.
\newblock Osiris: Hunting for integer bugs in ethereum smart contracts.
\newblock In {\em Proceedings of the 34th annual computer security applications
  conference}, pages 664--676, 2018.

\bibitem{wu2022TokenScout}
C.~Wu, J.~Chen, Z.~Zhao, K.~He, G.~Xu, Y.~Wu, H.~Wang, H.~Li, Y.~Liu, and
  Y.~Xiang.
\newblock Tokenscout: Early detection of ethereum scam tokens via temporal
  graph learning.
\newblock In {\em ACM CCS}, 2022.

\bibitem{xu2023lmpa}
X.~Xu, Z.~Zhang, S.~Feng, Y.~Ye, Z.~Su, N.~Jiang, S.~Cheng, L.~Tan, and
  X.~Zhang.
\newblock Lmpa: Improving decompilation by synergy of large language model and
  program analysis.
\newblock {\em arXiv:2306.02546}, 2023.

\bibitem{yu2021ponzi}
S.~Yu, J.~Jin, Y.~Xie, J.~Shen, and Q.~Xuan.
\newblock Ponzi scheme detection in ethereum transaction network.
\newblock In {\em Blockchain and Trustworthy Systems (BlockSys)}, 2021.

\bibitem{zhang2020ethploit}
Q.~Zhang, Y.~Wang, J.~Li, and S.~Ma.
\newblock Ethploit: From fuzzing to efficient exploit generation against smart
  contracts.
\newblock In {\em IEEE International Conference on Software Analysis, Evolution
  and Reengineering (SANER)}, 2020.

\bibitem{zheng2023securing}
Z.~Zheng, W.~Chen, Z.~Zhong, Z.~Chen, and Y.~Lu.
\newblock Securing the ethereum from smart ponzi schemes: Identification using
  static features.
\newblock {\em ACM Transactions on Software Engineering and Methodology}, 2023.

\end{thebibliography}

\end{document}